title page

# A Dialogue-Based Framework for Correcting Multimodal Errors in AI-Assisted STEM Education

Akshay Syal, Lawrence Swaminathan Xavier Prince, Evin Gultepe, Nik Bear Brown, Srinivas Sridhar

## Abstract

Large Language Models (LLMs) are democratizing access to personalized tutoring; however, their effectiveness is hindered by challenges in processing multimodal content, which limits AI's potential to provide equitable, high-quality STEM support.

This study evaluates LLM performance on multimodal physics problems, identifies specific failure modes through empirical error taxonomy, and tests practical interventions designed to overcome multimodal processing limitations. We assessed three publicly available LLMs (Claude, Gemini, and ChatGPT) on multimodal physics problems from the OpenStax database and compared the results with text-only performance. An empirically derived error taxonomy was developed through pilot testing, followed by evaluation of a structured multimodal dialogue intervention.

All three models achieved near-ceiling accuracy (96%) on text-only physics problems. Performance declined substantially on multimodal problems, consistent with what we term the Multimodal Interference Effect. Error analysis identified four failure modes: visual processing errors, context misinterpretation, mathematical computational errors, and hybrid errors, with visual processing errors being the most prevalent. The structured dialogue intervention corrected 82% of errors overall; visual processing errors were corrected at 100% across all models.

Educators and students can implement these interventions immediately, requiring no model retraining, to improve AI tutoring reliability on image-rich STEM content, advancing equitable access to high-quality learning support.

## 1. Introduction

Just as the internet democratized access to information, artificial intelligence is now democratizing access to personalized instruction. Large Language Models (LLMs) offer the potential to provide every student with an intelligent tutor capable of explaining concepts, answering questions, and guiding problem-solving(Adel et al., 2024; Purba et al., 2025; Xu et al., 2024). These opportunities were historically limited to those with access to expert teachers, private tutors, or well-resourced institutions. This technological shift represents a fundamental opportunity to address educational inequities by making high-quality, adaptive learning support universally accessible.

In STEM education, particularly, where personalized guidance is crucial for developing problem-solving skills and conceptual understanding, AI tutors could transform how millions of students worldwide learn physics, chemistry, biology, and mathematics. Early evidence suggests that AI-assisted learning can improve student outcomes, increase engagement, and provide support at scales impossible with traditional one-to-one tutoring(Kestin et al., 2025; Li et al., 2025; Sun et al., 2025). However, realizing this potential requires that AI systems can handle the full complexity of STEM content, including the multimodal representations central to these disciplines.

STEM education inherently relies on multimodal representations. Physics students must interpret velocity-time graphs and free-body diagrams. Chemistry students analyze molecular structures and reaction mechanisms. Biology students examine cell diagrams and anatomical images. Engineering students interpret circuit schematics, thermodynamic cycle diagrams, or signal waveforms while solving design problems. These visual representations are not supplementary; they are fundamental to understanding and communicating scientific knowledge. A tutoring system that cannot reliably process and reason for such multimodal content remains limited in its educational utility.

Current LLMs, despite remarkable text-processing capabilities, demonstrate inconsistent performance when problems integrate visual and textual information(Cao et al., 2024; Fu et al., 2025). This multimodal processing gap creates a barrier to equitable AI-assisted learning: students working on image-rich problems, which often require expert guidance, may receive unreliable or incorrect support. Understanding and addressing these limitations is essential for fulfilling AI’s promise of democratized educational access.

While recent studies have begun documenting LLM performance on multimodal STEM problems through systematic benchmarking efforts, a critical gap remains in understanding how to overcome identified limitations in practice. Existing research has primarily followed two paths. First, characterization-focused studies establish specialized benchmarks to evaluate model capabilities(Fu et al., 2025; Jia et al., 2025; Jian et al., 2025). These benchmarks document where and how models fail, providing valuable diagnostic information but stopping short of testing remediation strategies.

Second, solution-focused approaches develop specialized systems through model fine-tuning on domain-specific data(Azerbayev et al., 2023; Liu et al., 2023). While these specialized approaches demonstrate performance improvements, they require extensive computational resources for model training (often thousands of GPU hours), large domain-specific datasets, machine learning expertise for implementation, or subscription-based access that may limit widespread availability to diverse learners and under-resourced institutions.

These approaches, whether benchmark-focused or specialization-focused, leave unaddressed the practical question of how educators and students can improve general-purpose, publicly accessible LLM performance on multimodal STEM content through actionable intervention strategies. Without actionable strategies to address known limitations, the growing body of knowledge about AI shortcomings risks simply documenting barriers rather than dismantling them, potentially perpetuating educational inequities where students working on image-rich problems receive less reliable support than those on text-based problems.

Recent work has begun exploring intervention strategies to enhance multimodal reasoning performance. Zhang et al. (2023) demonstrated that separating rationale generation from answer inference through a two-stage framework substantially improves accuracy on science questions(Zhang et al., 2023). Similarly, Anand et al. (2024) incorporated Multi-Image Chain-of-Thought prompting alongside model fine-tuning(Anand et al., 2024). While these studies establish that strategic prompting and structured reasoning can enhance multimodal problem-solving, both approaches ultimately relied on specialized model training which raises a fundamental question for educational accessibility: Can the benefits of structured reasoning and strategic prompting be realized through intervention strategies alone, applied to general-purpose publicly-available models, without requiring specialized training? Answering this question is essential for democratizing improved AI tutoring, as prompt-based interventions would be immediately accessible to any educator or student, whereas fine-tuning remains confined to those with substantial computational resources.

This study addresses this question by isolating and systematically testing prompting-based intervention strategies designed to overcome specific multimodal failure modes. We developed practical intervention strategies designed to overcome failure modes specific to multimodal questions. Our contribution is threefold: empirically deriving an error taxonomy through pilot-grounded analysis of actual LLM failures on authentic physics problems; designing targeted interventions matched to visual error categories; and systematically evaluating whether these interventions significantly improve performance, thereby expanding the range of problems for which AI can provide reliable educational support. By focusing on actionable approaches implementable by educators, platform developers, or students themselves, this work bridges the gap between understanding AI limitations and overcoming them in practice.

From an instructional design perspective, the structured multimodal dialogue developed in this study functions as a scaffolding strategy that decomposes a complex, visually integrated task into discrete, sequenced steps. This approach draws on the principles of cognitive load management, where separating the perceptual and conceptual demands of a task reduces the likelihood of processing failures at either stage. Rather than presenting the image and the problem simultaneously and expecting the model to integrate both, the dialogue first elicits an explicit visual description, corrects any misidentified elements, and only then prompts problem-solving. In doing so, it mirrors pedagogical scaffolding techniques used in human tutoring contexts, where learners are guided to encode relevant information before applying domain knowledge. The practical significance of this framing is that the intervention is not merely a prompting workaround, but it is a replicable instructional pattern that educators, platform developers, and students can adopt immediately, without requiring changes to the underlying model or access to specialized computational resources. This approach positions structured multimodal dialogue as a contribution to the growing body of research on AI-assisted STEM instruction, offering a concrete, evidence-based strategy for extending the reliability of general-purpose LLMs into the image-rich problem spaces that define STEM education.

# 2. Methods

## 2.1 Question Set Development

Questions were drawn from the OpenStax Physics database(OpenStax), a freely available, peer-reviewed educational resource widely used in undergraduate physics instruction. We selected 100 text-based questions as a baseline and 44 multimodal questions for the main evaluation. Each multimodal problem included graphs, diagrams, tables, or visual equations essential to the solution, spanning quantitative calculation, conceptual reasoning, and graph interpretation. Problems requiring graphical outputs were excluded due to the current LLM drawing limitations. The 44 multimodal questions represent the eligible problems we could identify in the OpenStax database that meet these inclusion criteria.

Answer keys were drawn from OpenStax; for non-straightforward problems, a physics PhD reviewed solutions to establish definitive correct responses. All questions are publicly accessible, with chapter and problem references in supplementary materials.

## 2.2 LLM Selection and Educational Context

Three widely accessible LLMs were evaluated: GPT-5, Claude Sonnet-4.5, and Gemini-3 Pro Preview. These models were selected to balance accessibility with task-specific capability alignment in STEM problem-solving contexts. For text-only baseline evaluation, reasoning-optimized variants (e.g., Claude Opus 4.1) were used to isolate core physics reasoning ability independent of visual inputs. For multimodal evaluation, vision-integrated model configurations (e.g., Claude Sonnet-4.5 and Gemini-3 Pro Preview) were employed to assess each model's ability to jointly process visual and textual information, which is essential for image-dependent physics problems. Although reasoning-optimized models also support multimodal inputs, we used vision-integrated variants to better reflect how these systems are typically deployed for multimodal tasks in practice. Within this framework, Claude models were included for their strong performance in structured reasoning and step-by-step analytical tasks, Gemini for its native multimodal architecture and strength in vision-language understanding, and GPT-5 as a widely adopted general-purpose model with balanced reasoning and multimodal capabilities representative of typical user-facing systems. This design also reflects real-world deployment patterns in widely used AI systems, where model selection is often dynamically adjusted based on task requirements such as modality and reasoning complexity, while still maintaining controlled evaluation conditions. All models were accessed between 09/2025 and 02/2026; exact version identifiers are documented in Appendix B.

## 2.3 Evaluating Baseline Performance

For text-only problems, models were tested via their respective web interfaces using a consistent prompt designed to reflect how a student might approach homework: *"I will ask you to solve physics questions. You must solve them and then give me the final answer at the end. Are you ready?"* Answers were manually verified against the OpenStax solution key.

For multimodal problems, each problem image was encoded into an API-compatible format and submitted with the following prompt: *"You are an expert physics professor. Solve the problem and all sub-parts step-by-step. For each part: write the key equation, then substitute numbers, then compute. State assumptions if needed. Round your calculations to 2 decimal places. Be precise and concise — no extra commentary."* The model state was reset between questions to prevent cross-question leakage. Correctness was assessed using an independent multi-model LLM-based evaluator (see Appendix A).

Models were accessed through a combination of web interface and programmatic workflows due to differences in platform availability and version stability during the study period. Certain model versions were more reliably accessible through their native web interfaces, while others were evaluated using API-compatible pipelines to enable controlled input formatting and reproducible execution. Programmatic evaluation was designed to closely simulate interactive chat-based behavior by maintaining structured prompts and resetting model state between questions, thereby approximating the functional conditions of web-based usage. While interface-level differences (e.g., system prompts or context handling) may introduce minor variability, prior to work and internal validation suggest that such effects are limited when prompts and state management are standardized. As a result, we treat performance differences as primarily reflective of model capabilities rather than interface artifacts.

## 2.4 Error Taxonomy Development: Understanding Failure Modes

All incorrect responses were analyzed thematically, yielding four error categories:

- **Context Misinterpretation (C):** Incorrect assumptions about problem approaches or application of wrong physics concepts.
- **Visual Processing Errors (V):** Failures to accurately extract information from graphs, diagrams, or images.
- **Mathematical Computation Errors (M):** Incorrect calculations, unit errors, or computational logic failures.
- **Hybrid Errors (H):** Combinations of the above.

## 2.5 Testing Interventions - Structured Multimodal Dialogue

The intervention followed a three-step dialogue: (i) *"Describe what you observe in the figure/graph,"* (ii) researcher correction of any misidentified visual elements, and (iii) *"Solve the problem"* (or *"Read the question again and solve the problem"* when the description was lengthy).  In step (ii), corrections were restricted to observable visual properties in order to align the model's perception with the source image. For instance, researcher input was limited to clarifying the direction of a force vector, or the numerical value of a graph's intercept without providing any hints regarding the underlying physics principles or formulas.

For GPT-5 and Gemini-3-pro, interventions were executed programmatically with full chat history maintained per question; for Sonnet-4.5, interventions were conducted manually via the web interface with no carry-over between questions.

Given the elevated rate of context misinterpretation errors in multimodal questions (20%) relative to text-only questions (3%), we also evaluated whether the intervention could address context errors co-occurring with visual processing challenges.

### 2.6 Statistical Analysis

Model performance was compared using Cochran's Q test for paired proportions. To analyze differences in error distributions, a Chi-square test with Monte Carlo permutation (100,000 iterations) was employed. This non-parametric approach was selected to provide an exact p-value, accounting for the small sample size and the presence of expected cell frequencies below 5.  Success rates were reported with 95% exact confidence intervals (CI) calculated using the Clopper-Pearson method, $\alpha = 0.05$. This conservative approach was chosen to ensure reliable coverage, given the small sample size.

## 3. Results

### 3.1 Baseline Performance: Current Capabilities and Limitations

To establish a baseline for the user-friendly, public LLM models, we first tested a 100-question database. On text-only physics problems (Table 1), all three models demonstrated strong performance with an average accuracy of 96% (95% CI [0.93, 0.98]). This high baseline performance on text-only questions established that the models possess underlying physics knowledge and reasoning capabilities, setting the stage for examining what changes when visual information becomes essential. Cochran's Q test revealed no statistically significant differences among the models ($Q = 3.6$, $df = 2$, $p = 0.17$), indicating that all three models possess similar baseline capabilities when visual processing is not required

*Table 1 Success rate on text-only physics problems (n = 100).*

| *Model* | *% Success* | *Correct* | *Incorrect* | *Context Misinterpretation* | *Computation Errors* |
|---|---|---|---|---|---|
| *Claude* | 94 | 94 | 6 | 3 | 3 |
| *Gemini* | 97 | 97 | 3 | 0 | 3 |
| *ChatGPT* | 97 | 97 | 3 | 1 | 2 |

| combined[1] | 96 | 288 | 12 | 4 | 8 |
|---|---|---|---|---|---|

However, on multimodal questions with a figure essential for solving the problem (Table 2), overall accuracy dropped to 74% (95% CI [0.66, 0.82]). The success difference between the text-based and the multimodal questions can be seen better in Figure 1. Unlike the text-only baseline, where performance was uniform, the multimodal task revealed a highly significant difference in model capabilities (Cochran's Q = 10.2, df = 2, p = 0.006). This indicates that the addition of visual data creates a performance gap that was not present in pure text-based reasoning.

*Table 2 Success rate on multimodal physics problems (n = 44).*

| *Model* | *% Success* | *Correct* | *Incorrect* | *Visual Processing* | *Context Misinterpretation* | *Hybrid* |
|---|---|---|---|---|---|---|
| *Claude* | 64 | 28 | 16 | 9 | 2 | 5 |
| *Gemini* | 89 | 39 | 5 | 3 | 1 | 1 |
| *ChatGPT* | 70 | 31 | 13 | 5 | 6 | 2 |
| *combined[2]* | 74 | 98 | 34 | 17 | 8 | 8 |

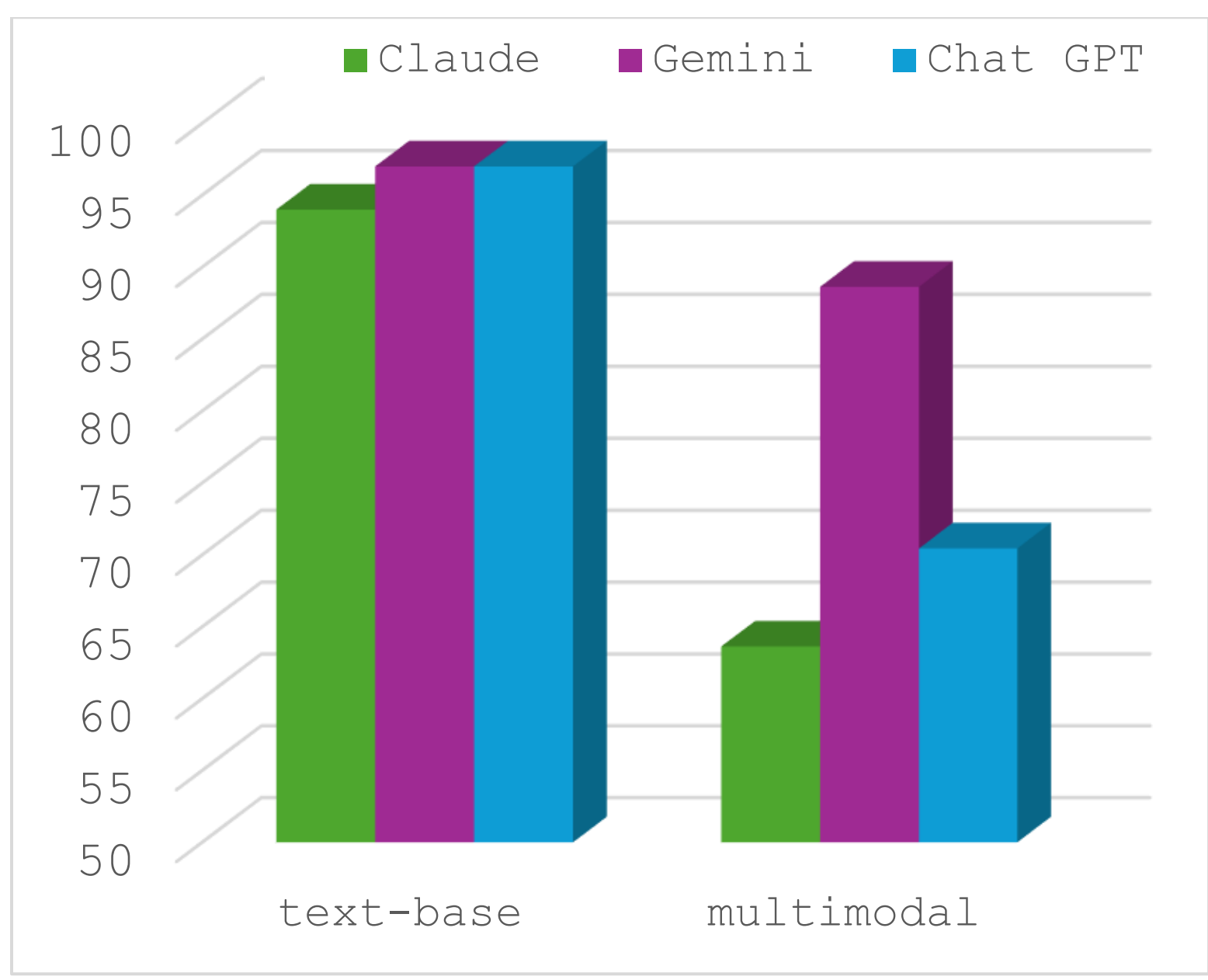

*Figure 1 Percent success of the baseline performance of the publicly available models on text-based or multimodal physics questions.*

Careful examination of the solutions that resulted in wrong answers gave us four main error categories: Context Misinterpretation, Mathematical Computation, Visual Processing, and Hybrid, which can be multiple combinations of these errors. Error category distribution can be seen in Figure 2. It is worth noting that Computation errors appeared only in text-only problems in our sample. In contrast, Visual and Hybrid errors appeared only in multimodal problems, suggesting that task modality may shape the type of reasoning failures that emerge. This pattern suggests that error type may be a function of task modality rather than model architecture alone, providing a rationale for interventions targeted specifically at visual processing failures.

---

[1] Combined row reflects totals across all three models (N = 300)

[2] Combined row reflects totals across all three models (N = 132)

For multi-modal problem-solving, descriptive results suggested that Claude was more prone to visual processing errors, while ChatGPT exhibited a higher frequency of context misinterpretation. However, a Monte Carlo permutation test (100,000 iterations) confirmed that these differences in error type distribution across models were not statistically significant $\chi^2 = 4.53$, df = 4, $p = 0.364$). Subsequent pairwise comparisons also yielded no significant results: Claude vs. ChatGPT $p = 0.152$; Claude vs. Gemini $p = 1.000$; Gemini vs. ChatGPT $p = 0.792$, suggesting a relatively uniform error profile across the tested architectures.

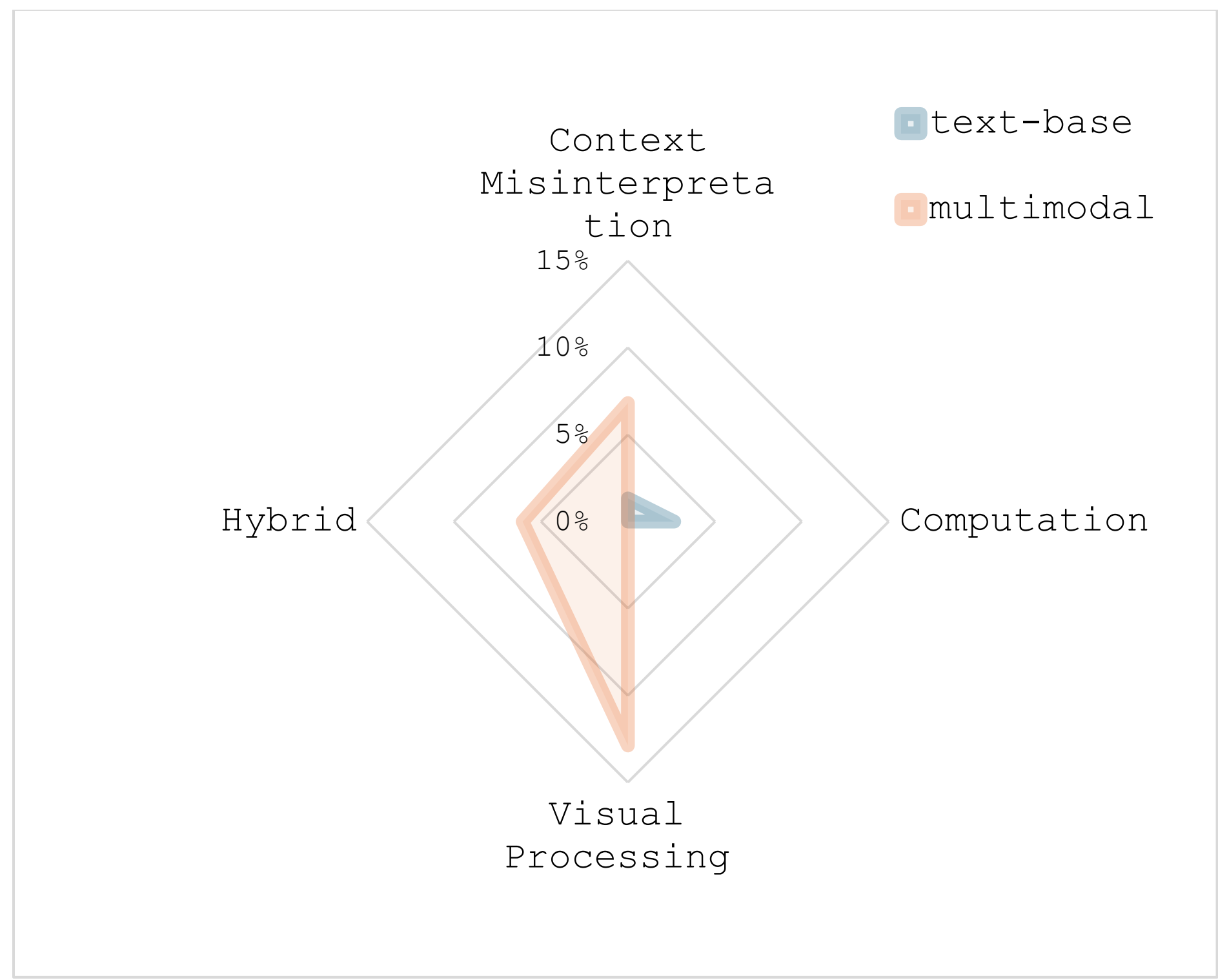

*Figure 2 Distribution of error categories, expressed as a percentage of total errors, for text-based and multimodal physics questions averaged over the tested models.*

## 3.2 Intervention Effectiveness: Pathways to Expanded Capability

The structured dialogue intervention was applied to all incorrect baseline responses, testing its effectiveness across different error categories. Overall, the intervention technique successfully corrected 28 of 34 baseline errors (82% success rate, 95% CI [0.655, 0.932]) across the three models, demonstrating substantial effectiveness in improving multimodal physics problem-solving without requiring model fine-tuning or specialized training ( Table 3). Visual processing errors, the most common failure mode, were corrected at 100% across all models (95% CI [0.815, 1.000]).

Following the intervention, overall multimodal accuracy across all three models rose to 95% (95% CI: [0.904, 0.983]) (Figure 3).

*Table 3 The effectiveness of the structured multimodal dialogue intervention by model and error type*

| Model | Total Errors Fixed | Visual Processing | Context Misinterpretation | Hybrid |
|---|---|---|---|---|
| Claude | 13/16 (81%) | 9/9 (100%) | 0/2 (0%) | 4/5 (80%) |
| Gemini | 4/5 (80%) | 3/3 (100%) | 0/1 (0%) | 1/1 (100%) |
| Chat GPT | 11/13 (85%) | 5/5 (100%) | 4/6 (67%) | 2/2 (100%) |
| combined | 28/34 (82%) | 17/17 (100%) | 4/9 (44%) | 7/8 (88%) |

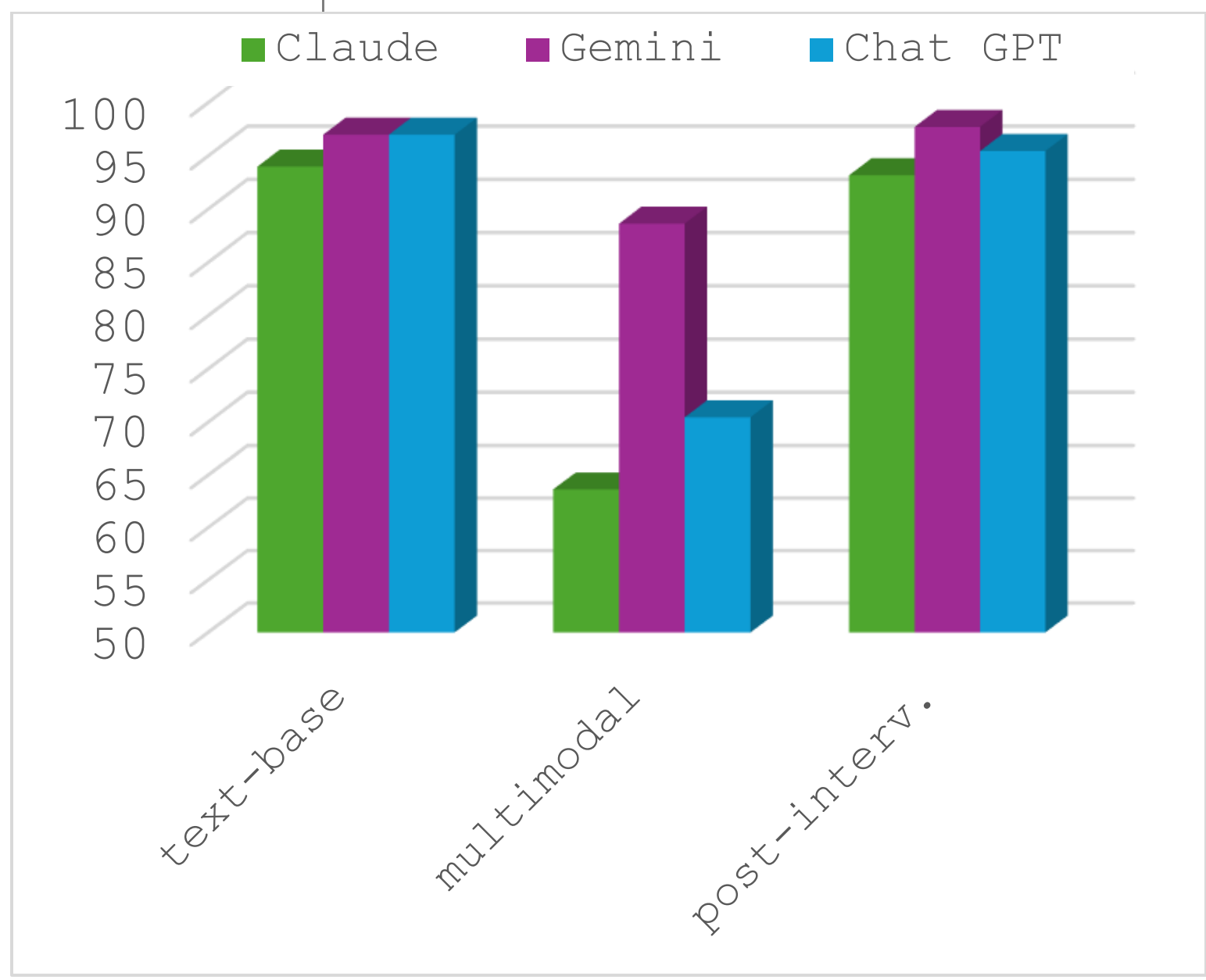

*Figure 3 With the structured dialogue intervention, the percent success of multimodal performance reaches closer to text-based performance for all the models tried.*

# 4. Discussion

## 4.1 Principal Findings

The substantial performance decline on multimodal problems, from a mean of 96% on text-only questions to between 64% and 89% depending on the model, points to a universal multimodal processing cost. We term this the *Multimodal Interference Effect*: while all three models demonstrate high proficiency in pure-text physics reasoning, the requirement to map visual representations to physical concepts introduces a layer of processing complexity that degrades performance even on aspects of problems that are not inherently visual. The rise in context misinterpretation errors in the multimodal condition (Table 2) relative to the text-only baseline (Table 1) supports this interpretation, i.e., visual processing appears to consume attentional or representational resources that would otherwise sustain accurate problem framing.

The intervention results further illuminate the structure of this interference. The 100% correction rate for visual processing errors demonstrates that models possess the underlying physics reasoning necessary to solve these problems correctly; the barrier lies specifically in visual information extraction, not in domain knowledge. Once accurate visual descriptions are provided, models reliably succeed. By contrast, context misinterpretation errors proved more resistant, corrected in only 44% of cases overall and not at all for Claude and Gemini. This differential suggests that context errors in the multimodal setting have a distinct etiology: they may reflect a deeper miscalibration of problem framing induced by the presence of visual input, one that explicit image description alone is insufficient to resolve.

## 4.2 Implications for STEM Education

These findings have direct implications for realizing AI's potential to democratize access to quality STEM education. The baseline results provide educators with a calibrated picture of where current AI tutors are reliable and where they are not. The intervention results offer a more optimistic counterpoint: current multimodal barriers are largely addressable without specialized resources. Structured dialogues that guide models through explicit visual processing before problem-solving can substantially close the performance gap visible in Figure 3, bringing multimodal accuracy in line with text-based performance for most problem types. These strategies can be immediately implemented by platform developers, instructional designers, or students themselves, requiring no access to model weights, training data, or computational infrastructure.

The multimodal processing challenges identified here extend beyond physics problems. Chemistry students analyzing molecular structures or spectroscopy data, and biology students examining cell diagrams or phylogenetic trees, face comparable demands. Our error taxonomy and intervention method are transferable to these domains.

## 4.3 From Research to Practice: Implementation Pathways

Our findings suggest three complementary pathways for improving AI-assisted learning on multimodal STEM content, each suited to different stakeholders and implementation contexts.

For educators: The error taxonomy developed here enables more informed integration of AI tutoring into instruction. By understanding which problem types are prone to visual or context errors, educators can design curricula that leverage AI support where it is reliable while ensuring human scaffolding remains available for content where it is not. Educators can also teach students the prompting literacy needed to elicit better AI performance, for instance, explicitly requesting step-by-step visual descriptions before problem-solving, as an increasingly important competency in AI-mediated learning environments.

For students: Even without any system-level changes, understanding the structured dialogue intervention empowers more effective use of available AI tools. A student who knows to ask "describe what you see in this graph" before "solve this problem" is better positioned to obtain reliable guidance than one who submits both simultaneously. This kind of informed interaction strategy can meaningfully improve learning outcomes without requiring any additional resources.

For platform developers: Integrating structured multimodal dialogue as a default interaction pattern for image-containing problems represents a low-cost, high-impact design intervention. Automating the visual description and correction steps within a tutoring interface could extend the benefits observed here to all users without requiring individual prompting expertise.

### 4.4 Limitations and Future Directions

This study has several limitations that should inform interpretation and guide future work. The 44-question multimodal sample reflects the eligible problems identified in the OpenStax database under our inclusion criteria rather than a statistically predetermined sample size, and additional eligible problems may exist within the database that were not captured. Future work drawing from larger or multi-textbook databases would strengthen generalizability. Additionally, binary correctness scoring does not capture partial understanding with pedagogical value, i.e., a response that correctly identifies the relevant physics principle but makes a computational error is treated identically to a completely incorrect response, which may underestimate the educational utility of AI tutoring on multimodal content.

We note that using different model variants across conditions introduces a degree of heterogeneity that would not arise in a fully controlled single-model evaluation. However, since our primary research question concerns the effect of the structured dialogue intervention rather than direct between-model performance comparisons, this design choice does not affect the validity of the core intervention finding. Although this design introduces heterogeneity that precludes direct cross-model comparisons, the core intervention finding, i.e., that structured dialogue substantially reduces visual processing errors, was replicated across all three model families, supporting the generalizability of this specific result.

Regarding the intervention protocol, corrections in step (ii) were constrained to observable visual properties of the source image; however, no inter-rater verification was performed to confirm that the researcher's input remained free of domain-relevant information. Future work should validate this constraint through blind auditing of correction content and explore whether correction prompts can be generalized or automated, which would reduce reliance on researcher judgment and strengthen the intervention's practical accessibility.

While the 17/17 correction rate for visual processing errors in this work is striking, the Clopper-Pearson 95% CI [0.815, 1.000] reflects considerable uncertainty. A more definitive conclusion needs replication with larger and more diverse problem sets.

Finally, the evaluated model versions will continue to evolve, and findings specific to current architectures may not hold as multimodal capabilities improve. The intervention was also tested on each strategy independently rather than in combination, and problems requiring graphical outputs were excluded due to current LLM limitations. Most importantly, all testing was conducted under controlled conditions; how students interact with AI tutors in authentic learning environments with varying prompting literacy, subject confidence, and task framing remains the critical open question for translating these findings into real educational impact.

## 5. Conclusions

Three widely accessible LLMs solve text-based physics problems with approximately 96% accuracy, but this accuracy declined by as much as 30% depending on the model when problems require the interpretation of diagrams, graphs, or other visual elements. This gap is driven primarily by visual processing errors and by an elevated rate of context misinterpretation that appears to co-occur with the added cognitive load of visual input, a pattern we term Multimodal Interference Effect.

A simple, three-step structured dialogue intervention, eliciting an explicit visual description, correcting misidentified elements, and then prompting problem-solving, achieved an 82% overall error correction rate without requiring any model fine-tuning, specialized training data, or computational resources beyond standard API access. Visual processing errors were resolved at a rate of 100% across all models. This demonstrates that a significant portion of multimodal failure is not intractable; it arises from insufficient engagement with visual content at the point of inference and can be corrected through structured prompting alone. Because these interventions require no specialized resources or model retraining, they are immediately accessible to the under-resourced institutions and individual learners who stand to benefit most from equitable AI tutoring support.

The multimodal barrier is real and measurable, but largely addressable through prompting strategies immediately available to any educator, student, or platform developer. The framework and methodology developed here provide a replicable foundation for ongoing evaluation as LLM capabilities evolve.

## Funding

Open access publishing of this article was supported by Northeastern University Library through its Read and Publish Agreement with Springer Nature.

## Competing Interests

The authors declare no competing interests.

# Supplementary Materials

## Appendix A: List of Problems

All questions were taken from OpenStax University Physics Volume I, Unit 1: Mechanics:

| Question_ID | Question |
|---|---|
| Chapter#3Question#21 | The severity of a fall depends on your speed when you strike the ground. All factors but the acceleration from gravity being the same, how many times higher could a safe fall occur on the Moon than on Earth (gravitational acceleration on the Moon is about one-sixth that of the Earth)? |
| Chapter#3Question#25 | A car is 2.0 km west of a traffic light at t = 0 and 5.0 km east of the light at t = 6.0 min. Assume the origin of the coordinate system is the light and the positive x direction is eastward. (a) What are the car's position vectors at these two times? (b) What is the car's displacement between 0 min and 6.0 min? |
| Chapter#3Question#27 | The position of a particle moving along the x-axis is given by m. (a) At what time does the particle cross the origin? (b) What is the displacement of the particle between t=3s and t = 6s |
| Chapter#3Question#29 | 29. On February 15, 2013, a superbolide meteor (brighter than the Sun) entered Earth's atmosphere over Chelyabinsk, Russia, and exploded at an altitude of 23.5 km. Eyewitnesses could feel the intense heat from the fireball, and the blast wave from the explosion blew out windows in buildings. The blast wave took approximately 2 minutes 30 seconds to reach ground |
| Chapter#3Question#35 | A particle moves along the x-axis according to x(t) = 10t - 2t^2m. (a) What is the instantaneous velocity at t = 2 s and t = 3 s? (b) What is the instantaneous speed at these times? (c) What is the average velocity between t = 2 s and t = 3 s? |
| Chapter#3Question#37 | A cheetah can accelerate from rest to a speed of 30.0 m/s in 7.00 s. What is its acceleration? |
| Chapter#3Question#41 | Assume an intercontinental ballistic missile goes from rest to a suborbital speed of 6.50 km/s in 60.0 s (the actual speed and time are classified). What is its average acceleration in |
| | meters per second squared and in multiples of g (9.80 m/s2 )? |
| Chapter#3Question#43 | A particle moves in a straight line at a constant velocity of 30 m/s. What is its displacement between t = 0 and t = 5.0 s? |
| Chapter#3Question#45 | A particle moves in a straight line with an initial velocity of 30 m/s and constant acceleration 30 m/s2 . (a) What is its displacement at t = 5 s? (b) What is its velocity at this same time? |
| Chapter#3Question#49 | At t = 10 s, a particle is moving from left to right with a speed of 5.0 m/s. At t = 20 s, the particle is moving right to left with a speed of 8.0 m/s. Assuming the particle's acceleration is constant, determine (a) its acceleration, (b) its initial velocity, and (c) the instant when its velocity is zero. |
| Chapter#3Question#51 | A bullet in a gun is accelerated from the firing chamber to the end of the barrel at an average rate of 6.2 x 10^5 m/s^2 for 8.10 x 10^-4 seconds. What is its muzzle velocity (that is, its final velocity)? |
| Chapter#3Question#57 | A powerful motorcycle can accelerate from rest to 26.8 m/s (100 km/h) in only 3.90 s. (a) What is its average acceleration? (b) Assuming constant acceleration, how far does it travel in that time? |
| Chapter#3Question#59 | A fireworks shell is accelerated from rest to a velocity of 65.0 m/s over a distance of 0.250 m. (a) Calculate the acceleration. (b) How long did the acceleration last? |
| Chapter#3Question#61 | A woodpecker's brain is specially protected from large accelerations by tendon-like attachments inside the skull. While pecking on a tree, the woodpecker's head comes to a stop from an initial velocity of 0.600 m/s in a distance of only 2.00 mm. (a) Find the acceleration in meters per second squared and in multiples of g, where g = 9.80 m/s2 . (b) Calculate the stopping time. (c) The tendons cradling the brain stretch, making its stopping distance 4.50 mm (greater than the head and, hence, less acceleration of the brain). What is the brain's acceleration, expressed in multiples of g? |
| Chapter#3Question#63 | A care package is dropped out of a cargo plane and lands in the forest. If we assume the care package speed on impact is 54 m/s (123 mph), then what is its acceleration? Assume the trees and snow stops it over a distance of 3.0 m. |
| Chapter#3Question#67 | Calculate the displacement and velocity at times of (a) 0.500 s, (b) 1.00 s, (c) 1.50 s, (d) 2.00 s, and (e) 2.50 s for a rock thrown straight down with an initial velocity of 14.0 m/s from the Verrazano Narrows Bridge in New York City. The roadway of this bridge is 70.0 m above the water. |
| Chapter#3Question#69 | A rescue helicopter is hovering over a person whose boat has sunk. One of the rescuers throws a life preserver straight down to the victim with an initial velocity of 1.40 m/s and observes that it takes 1.8 s to reach the water. (a) List the knowns in this problem. (b) How high above the water was the preserver released? Note that the downdraft of the helicopter reduces the effects of air resistance on the falling life preserver, so that an acceleration equal to that of gravity is reasonable. |
| Chapter#3Question#71 | A diver bounces straight up from a diving board, avoiding the diving board on the way down, and falls feet first into a pool. She starts with a velocity of 4.00 m/s and her takeoff point is 1.80 m above the pool. (a) What is her highest point above the board? (b) How long a time are her feet in the air? (c) What is her velocity when her feet hit the water? |
| Chapter#3Question#73 | A very strong, but inept, shot putter puts the shot straight up vertically with an initial velocity of 11.0 m/s. How long a time does he have to get out of the way if the shot was released at a height of 2.20 m and he is 1.80 m tall? |
| Chapter#3Question#75 | A kangaroo can jump over an object 2.50 m high. (a) Considering just its vertical motion, calculate its vertical speed when it leaves the ground. (b) How long a time is it in the air? |
| Chapter#3Question#77 | There is a 250-m-high cliff at Half Dome in Yosemite National Park in California. Suppose a boulder breaks loose from the top of this cliff. (a) How fast will it be going when it strikes the ground? (b) Assuming a reaction time of 0.300 s, how long a time will a tourist at the bottom |
| | have to get out of the way after hearing the sound of the rock breaking loose (neglecting the height of the tourist, which would become negligible anyway if hit)? The speed of sound is 335.0 m/s on this day. |
| Chapter#3Question#79 | Between t = 0 and t = t0<br><br>, a rocket moves straight upward with an acceleration given by a(t) = A - B t^1/2 , where A and B are constants. (a) If x is in meters and t is in seconds, what are the units of A and B? (b) If the rocket starts from rest, how does the velocity vary between t = 0 and t = t0? (c) If its initial position is zero, what is the rocket's position as a function of time during this same time interval? |
| Chapter#3Question#81 | A particle at rest leaves the origin with its velocity increasing with time according to v(t) = 3.2t m/s. At 5.0 s, the particle's velocity starts decreasing according to [16.0 – 1.5(t – 5.0)] m/s. This decrease continues until t = 11.0 s, after which the particle's velocity remains constant at 7.0 m/s. (a) What is the acceleration of the particle as a function of time? (b) What is the position of the particle at t = 2.0 s, t = 7.0 s, and t = 12.0 s? |
| Chapter#3Question#83 | An airplane leaves Chicago and makes the 3000-km trip to Los Angeles in 5.0 h. A second plane leaves Chicago one-half hour later and arrives in Los Angeles at the same time. Compare the average velocities of the two planes. Ignore the curvature of Earth and the difference in altitude between the two cities. |
| Chapter#3Question#85 | An object has an acceleration of 1.2cm/s^2 At t = 4 seconds its velocity is -3.4cm/s Determine the object's velocities at t =1s and t = 6s |
| Chapter#3Question#87 | A particle moving at constant acceleration has velocities of 2.0 m/s at t =2.0 s and –7.6 m/s at t = 5.2 s. What is the acceleration of the particle? |
| Chapter#3Question#89 | An electron is moving in a straight line with a velocity of 4 x 1-&5m/s. It enters a region 5.0 cm long where it undergoes an acceleration of 6 x 10^12 m/s^2 along the same straight line. (a) What is the electron's velocity when it |
| | emerges from this region? b) How long does the electron take to cross the region? |
| Chapter#3Question#91 | A motorcycle that is slowing down uniformly covers 2.0 successive km in 80 s and 120 s, respectively. Calculate (a) the acceleration of the motorcycle and (b) its velocity at the beginning and end of the 2-km trip. |
| Chapter#3Question#93 | Two trains are moving at 30 m/s in opposite directions on the same track. The engineers see simultaneously that they are on a collision course and apply the brakes when they are 1000 m apart. Assuming both trains have the same acceleration, what must this acceleration be if the trains are to stop just short of colliding? |
| Chapter#3Question#95 | A police car waits in hiding slightly off the highway. A speeding car is spotted by the police car doing 40 m/s. At the instant the speeding car passes the police car, the police car accelerates from rest at 4 m/s2 to catch the speeding car. How long does it take the police car to catch the speeding car? |
| Chapter#3Question#97 | Unreasonable results A runner approaches the finish line and is 75 m away; her speed at this position is 8 m/s. She accelerates opposite to the motion at this point at 0.5 m/s2 . How long does it take her to cross the finish line from 75 m away? Is this reasonable? |
| Chapter#3Question#101 | A ball is thrown straight up. It passes a 2.00-m-high window 7.50 m off the ground on its path up and takes 1.30 s to go past the window. What was the ball’s initial velocity? |
| Chapter#3Question#103 | A soft tennis ball is dropped onto a hard floor from a height of 1.50 m and rebounds to a height of 1.10 m. (a) Calculate its velocity just before it strikes the floor. (b) Calculate its velocity just after it leaves the floor on its way back up. (c) Calculate its acceleration during contact with the floor if that contact lasts 3.50 ms (d) How much did the ball |
| | compress during its collision with the floor, assuming the floor is absolutely rigid? |
| Chapter#3Question#105 | Compare the time in the air of a basketball player who jumps 1.0 m vertically off the floor with that of a player who jumps 0.3 m vertically. |
| Chapter#3Question#107 | A hot-air balloon rises from ground level at a constant velocity of 3.0 m/s. One minute after liftoff, a sandbag is dropped accidentally from the balloon. Calculate (a) the time it takes for the sandbag to reach the ground and (b) the velocity of the sandbag when it hits the ground. |
| Chapter#3Question#109 | An object is dropped from a height of 75.0 m above ground level. (a) Determine the distance traveled during the first second. (b) Determine the final velocity at which the object hits the ground. (c) Determine the distance traveled during the last second of motion before hitting the ground. |
| Chapter#3Question#111 | An object is dropped from a roof of a building of height h. During the last second of its descent, it drops a distance h/3. Calculate the height of the building. |
| Chapter#3Question#113 | The position of a particle moving along the x-axis varies with time according to x(t) = 5t^2 - rt^3 m. Find (a) the velocity and acceleration of the particle as functions of time, (b) the velocity and acceleration at t = 2.0 s, (c) the time at which the position is a maximum, (d) the time at which the velocity is zero, and (e) the maximum position. |
| Chapter#3Question#115 | In 1967, New Zealander Burt Munro set the world record for an Indian motorcycle, on the Bonneville Salt Flats in Utah, of 295.38 km/h. The one-way course was 8.00 km long. Acceleration rates are often described by the time it takes to reach 96.0 km/h from rest. If this time was 4.00 s and Burt accelerated at this rate until he reached his maximum speed, how long did it take Burt to complete the course? |
| Chapter#4Question#17 | The coordinates of a particle in a rectangular coordinate system are (1.0, –4.0, 6.0). What is the position vector of the particle? |
| Chapter#4Question#19 | 19. The 18th hole at Pebble Beach Golf Course is a dogleg to the left of length 496.0 m. The fairway off the tee is taken to be the x direction. |

| | A golfer hits his tee shot a distance of 300.0 m, corresponding to a displacement r1=300mi and hits his second shot 189.0 m with a displacement r2 = 172mi + 80.3mjWhat is the final displacement of the golf ball from where it started? |
|---|---|
| Chapter#4Question#21 | A cyclist rides 5.0 km due east, then 10.0 km 20 degrees west of north. From this point she rides 8.0 km due west. What is the final displacement from where the cyclist started? |
| Chapter#4Question#23 | The position of a particle is r(t) = 4t^2i -3j + 2t^3km(a) What is the velocity of the particle at 0 s and at 1 s? (b) What is the average velocity between 0 s and 1 s? |
| Chapter#4Question#25 | The F-35B Lighting II is a short-takeoff and vertical landing fighter jet. If it does a vertical takeoff to 20.00-m height above the ground and then follows a flight path angled at 30 degrees with<br>respect to the ground for 20.00 km, what is the final displacement? |
| Chapter#4Question#29 | The position of a particle for t > 0 is given by r=3t^2i -7t^3j -5t&-2km (a) What is the velocity as a function of time? (b) What is the acceleration as a function of time? (c) What is the particle's velocity at t = 2.0 s? (d) What is its speed at t = 1.0 s and t = 3.0 s? (e) What is the average velocity between t = 1.0 s and t = 2.0 s? |
| Chapter#4Question#31 | A particle has a position function r(t) = cost(1t)I + sin(1t)j + tk, where the arguments of the cosine and sine functions are in radians. (a) What is the velocity vector? (b) What is the acceleration vector? |
| Chapter#4Question#33 | A bullet is shot horizontally from shoulder height (1.5 m) with an initial speed 200 m/s. (a) How much time elapses before the bullet hits the ground? (b) How far does the bullet travel horizontally? |
| Chapter#4Question#39 | 39. A projectile is launched at an angle of 30 degrees and lands 20 s later at the same height as it was<br>launched. (a) What is the initial speed of the projectile? (b) What is the maximum altitude? (c) What is the range? (d) Calculate the displacement from the point of launch to the |
| | position on its trajectory at 15 s. |
| Chapter#4Question#41 | At a particular instant, a hot air balloon is 100 m in the air and descending at a constant speed of 2.0 m/s. At this exact instant, a girl throws a ball horizontally, relative to herself, with an initial speed of 20 m/s. When she lands, where will she find the ball? Ignore air resistance. |
| Chapter#4Question#43 | An athlete can jump a distance of 8.0 m in the broad jump. What is the maximum distance the athlete can jump on the Moon, where the gravitational acceleration is one-sixth that of Earth? |
| Chapter#4Question#45 | A rock is thrown off a cliff at an angle of 53 degrees with respect to the horizontal. The cliff is 100 m high. The initial speed of the rock is 30 m/s. (a) How high above the edge of the cliff does the rock rise? (b) How far has it moved horizontally when it is at maximum altitude? (c) How long after the release does it hit the ground? (d) What is the range of the rock? (e) What are the horizontal and vertical positions of the rock relative to the edge of the cliff at t = 2.0 s, t = 4.0 s, and t = 6.0 s? |
| Chapter#4Question#47 | A golfer on a fairway is 70 m away from the green, which sits below the level of the fairway by 20 m. If the golfer hits the ball at an angle of with an initial speed of 20 m/s, how close to the green does she come? |
| Chapter#4Question#49 | An astronaut on Mars kicks a soccer ball at an angle of 45 degrees with an initial velocity of 15 m/s. If the acceleration of gravity on Mars is 3.7 m/s2<br><br>, (a) what is the range of the soccer kick on a flat surface? (b) What would be the range of the same kick on the Moon, where gravity is one-sixth that of Earth? |
| Chapter#4Question#51 | MIT’s robot cheetah can jump over obstacles 46 cm high and has speed of 12.0 km/h. (a) If the robot launches itself at an angle of 60 degrees at this speed, what is its maximum height? (b) What would the launch angle have to be to reach a height of 46 cm? |
| Chapter#4Question#53 | 53. Drew Brees of the New Orleans Saints can throw a football 23.0 m/s (50 mph). If he angles the throw at 10 degrees from the horizontal, what distance does it go if it is to be caught at the same elevation as it was thrown? |
| Chapter#4Question#55 | A soccer goal is 2.44 m high. A player kicks the ball at a distance 10 m from the goal at an angle of 25 degrees. The ball hits the crossbar at the top of the goal. What is the initial speed of the soccer ball? |
| Chapter#4Question#57 | 57. In 1999, Robbie Knievel was the first to jump the Grand Canyon on a motorcycle. At a narrow part of the canyon (69.0 m wide) and traveling 35.8 m/s off the takeoff ramp, he reached the other side. What was his launch angle? |
| Chapter#4Question#59 | Aaron Rodgers throws a football at 20.0 m/s to his wide receiver, who is running straight down the field at 9.4 m/s. If Aaron throws the football when the wide receiver is 10.0 m in front of him, (a) at what angle does Aaron have to launch the ball so the ball will be at the same height as the receiver when the receiver makes it to 20.0 m in front of Aaron? (b) Will the receiver be able to catch the ball? |
| Chapter#4Question#61 | A particle travels in a circle of radius 10 m at a constant speed of 20 m/s. What is the magnitude of the acceleration? |
| Chapter#4Question#63 | A fairground ride spins its occupants inside a flying saucer-shaped container. If the horizontal circular path the riders follow has an 8.00-m radius, at how many revolutions per minute are the riders subjected to a centripetal acceleration equal to that of gravity? |
| Chapter#4Question#67 | A fan is rotating at a constant 360.0 rev/min. What is the magnitude of the acceleration of a point on one of its blades 10.0 cm from the axis of rotation? |
| Chapter#4Question#69 | The coordinate axes of the reference frame remain parallel to those of S, as moves away from S at a constant velocity (a) If at time t = 0 the origins coincide, what is the position of the origin in the |
| | S frame as a function of time? (b) How is particle position for and as measured in S and respectively, related? (c) What is the relationship between particle velocities (d) How are accelerations related? |
| Chapter#4Question#71 | The velocity of a particle in reference frame A is The velocity of reference frame A with respect to reference frame B is and the velocity of reference frame B with respect to C is What is the velocity of the particle in reference frame C? |
| Chapter#4Question#73 | A seagull can fly at a velocity of 9.00 m/s in still air. (a) If it takes the bird 20.0 min to travel 6.00 km straight into an oncoming wind, what is the velocity of the wind? (b) If the bird turns around and flies with the wind, how long will it take the bird to return 6.00 km? |
| Chapter#4Question#75 | A boat can be rowed at 8.0 km/h in still water. (a) How much time is required to row 1.5 km downstream in a river moving 3.0 km/h relative to the shore? (b) How much time is required for the return trip? (c) In what direction must the boat be aimed to row straight across the river? (d) Suppose the river is 0.8 km wide. What is the velocity of the boat with respect to Earth and how much time is required to get to the opposite shore? (e) Suppose, instead, the boat is aimed straight across the river. How much time is required to get across and how far downstream is the boat when it reaches the opposite shore? |
| Chapter#4Question#77 | A cyclist traveling southeast along a road at 15 km/h feels a wind blowing from the southwest at 25 km/h. To a stationary observer, what are the speed and direction of the wind? |
| Chapter#4Question#79 | A Formula One race car is traveling at 89.0 m/s along a straight track enters a turn on the race track with radius of curvature of 200.0 m. What centripetal acceleration must the car have to stay on the track? |

| Chapter#4Question#81 | The driver of a car moving at 90.0 km/h presses down on the brake as the car enters a circular curve of radius 150.0 m. If the speed of the car is decreasing at a rate of 9.0 km/h each second, what is the magnitude of the acceleration of the car at the instant its speed is 60.0 km/h? |
|---|---|
| Chapter#4Question#83 | An elephant is located on Earth's surface at a latitude Calculate the centripetal acceleration of the elephant resulting from the rotation of Earth around its polar axis. Express your answer in terms of the radius of Earth, and time T for one rotation of Earth. Compare your answer with g for lambda = 40 degrees. |
| Chapter#4Question#85 | A propeller blade at rest starts to rotate from t = 0 s to t = 5.0 s with a tangential acceleration of the tip of the blade at 3.00 m/s^2. The tip of the blade is 1.5 m from the axis of rotation. At t = 5.0 s, what is the total acceleration of the tip of the blade? |
| Chapter#4Question#87 | A particle's centripetal acceleration is ac= 4m/s^2 at t = 0 s where it is on the x-axis and moving counterclockwise in the xy plane. It is executing uniform circular motion about an axis at a distance of 5.0 m. What is its velocity at t = 10 s? |
| Chapter#4Question#89 | A particle located initially at 1.5j + 4k m undergoes a displacement of 2.5i + 3.2j - 1.2k<br><br>What is the final position of the particle? |
| Chapter#4Question#91 | A spaceship is traveling at a constant velocity of v(t) = 250 I m/s when its rockets fire, giving it an acceleration of a(t) = 3i + 4k m/s^2. What is its velocity 5 s after the rockets fire? |
| Chapter#4Question#93 | A long jumper can jump a distance of 8.0 m when he takes off at an angle of 45 degrees with respect to the horizontal. Assuming he can jump with the same initial speed at all angles, how much distance does he lose by taking off at 30 degrees |
| Chapter#4Question#95 | A mountain biker encounters a jump on a race course that sends him into the air at 60 degrees to the horizontal. If he lands at a horizontal distance of 45.0 m and 20 m below his launch |
| | point,<br> what is his initial speed? |
| Chapter#4Question#97 | A geosynchronous satellite orbits Earth at a distance of 42,250.0 km and has a period of 1 day. What is the centripetal acceleration of the satellite? |
| Chapter#4Question#99 | 99. World's Longest Par 3. The tee of the world's longest par 3 sits atop South Africa's Hanglip Mountain at 400.0 m above the green and can only be reached by helicopter. The horizontal distance to the green is 359.0 m. Neglect air resistance and answer the following questions. (a) If a golfer launches a shot that is with respect to the horizontal, what initial velocity must she give the ball? (b) What is the time to reach the green? |
| Chapter#5Question#27 | Astronauts in orbit are apparently weightless. This means that a clever method of measuring the mass of astronauts is needed to monitor their mass gains or losses, and adjust their diet. One way to do this is to exert a known force on an astronaut and measure the acceleration produced. Suppose a net external force of 50.0 N is exerted, and an astronaut's acceleration is measured to be 0.893m/s^2. (a) Calculate her mass. (b) By exerting a force on the astronaut, the vehicle in which she orbits experiences an equal and opposite force. Use this knowledge to find an equation for the acceleration of the system (astronaut and spaceship) that would be measured by a nearby observer. (c) Discuss how this would affect the measurement of the astronaut's acceleration. Propose a method by which recoil of the vehicle is avoided. |
| Chapter#5Question#33 | A powerful motorcycle can produce an acceleration of 3.5m/s^2 while traveling at 90.0 km/h. At that speed, the forces resisting motion, including friction and air resistance, total 400.0 N. (Air resistance is analogous to air friction. It always opposes the motion of an object.) What is the magnitude of the force that motorcycle exerts backward on the ground to produce its |
| | acceleration if the mass of the motorcycle with rider is 245 kg? |
| Chapter#5Question#37 | A particle of mass 2.0 kg is acted on by a single force F = 18i N (a) What is the particle's acceleration? (b) If the particle starts at rest, how far does it travel in the first 5.0 s? |
| Chapter#5Question#41 | 41. The weight of an astronaut plus his space suit on the Moon is only 250 N. (a) How much does the suited astronaut weigh on Earth? (b) What is the mass on the Moon? On Earth? |
| Chapter#5Question#43 | A rocket sled accelerates at a rate of 49m/s^2 . Its passenger has a mass of 75.0 kg. (a) Calculate the horizontal component of the force the seat exerts against his body. Compare this with his weight using a ratio. (b) Calculate the direction and magnitude of the total force the seat exerts against his body. |
| Chapter#5Question#45 | A body of mass 2.00 kg is pushed straight upward by a 25.0 N vertical force. What is its acceleration? |
| Chapter#5Question#47 | 47. A body with a mass of 10.0 kg is assumed to be in Earth's gravitational field with g = 9.80 m/s^2. What is the net force on the body if there are no other external forces acting on the object? |
| Chapter#5Question#51 | (a) What net external force is exerted on a 1100.0-kg artillery shell fired from a battleship if the shell is accelerated at 2.3 x 10^4 m/s^2 (b) What is the magnitude of the force exerted on the ship by the artillery shell, and why? |
| Chapter#5Question#57 | A team of nine members each engage in a tug-of-war, pulling in opposite directions on a horizontal rope. Each of the first team's members has an average mass of 68 kg and exerts an average force of 1350 N horizontally on the ground as they pull on the rope. Each of the second team's members has an average mass of 73 kg and exerts an average force of 1365 N horizontally on the ground as they pull on the rope in the opposite direction. (a) What is magnitude of the acceleration of the two teams, |
|  | and which team wins? (b) What is the tension in the section of rope between the teams? |
| Chapter#5Question#83 | A force acts on a car of mass m so that the speed v of the car increases with position x as v=kx^2 , where k is constant and all quantities are in SI units. Find the force acting on the car as a function of position. |
| Chapter#5Question#85 | 85. Two boxes, A and B, are at rest. Box A is on level ground, while box B rests on an inclined plane tilted at angle teta with the horizontal. (a) Write expressions for the normal force acting on each block. (b) Compare the two forces; that is, tell which one is larger or whether they are equal in magnitude. (c) If the angle of incline is , 10 degrees which force is greater? |
| Chapter#5Question#89 | In building a house, carpenters use nails from a large box. The box is suspended from a spring twice during the day to measure the usage of nails. At the beginning of the day, the spring stretches 50 cm. At the end of the day, the spring stretches 30 cm. What fraction or percentage of the nails have been used? |
| Chapter#5Question#91 | Two forces are applied to a 5.0-kg object, and it accelerates at a rate of 2m/s^2 in the positive y-direction. If one of the forces acts in the positive x-direction with magnitude 12.0 N, find the magnitude of the other force. |
| Chapter#5Question#95 | 95. On June 25, 1983, shot-putter Udo Beyer of East Germany threw the 7.26-kg shot 22.22 m, which at that time was a world record. (a) If the shot was released at a height of 2.20 m with a projection angle of45 degrees , what was its initial velocity? (b) If while in Beyer's hand the shot was accelerated uniformly over a distance of 1.20 m, what was the net force on it? |
| Chapter#5Question#99 | A 120-kg astronaut is riding in a rocket sled that is sliding along an inclined plane. The sled has a horizontal component of acceleration of and a downward component of 3.8m/s^2 . Calculate the magnitude of the force on the rider by the sled. (Hint: Remember that gravitational acceleration must be considered.) |

| Chapter#5Question#101 | Suppose that you are viewing a soccer game from a helicopter above the playing field. Two soccer players simultaneously kick a stationary soccer ball on the flat field; the soccer ball has mass 0.420 kg. The first player kicks with force 162 N at 9 degrees north of west. At the same instant, the second player kicks with force 215 N at 15 degrees east of south. Find the acceleration of the ball in and form. |
|---|---|
| Chapter#5Question#103 | A 0.0502-kg pair of fuzzy dice is attached to the rearview mirror of a car by a short string. The car accelerates at constant rate, and the dice hang at an angle of 3.20 degrees from the vertical because of the car's acceleration. What is the magnitude of the acceleration of the car? |
| Chapter#5Question#107 | F1 = (-3.00 i + 2.00 j) N<br>F2 = (6.00 i - 4.00 j) N<br>F3 = (2.00 i + 5.00 j) N<br><br>Acceleration magnitude = 4.23 m/s^2<br><br>a) Find acceleration vector in terms of m<br>b) Find mass<br>c) From rest, find speed after 5.00 s<br>d) Find velocity components after 5.00 s |
| Chapter#5Question#109 | A drone is being directed across a frictionless ice-covered lake. The mass of the drone is 1.50 kg, and its velocity is 3.00i m/s . After 10.0 s, the velocity is 9i + 4j m/s . If a constant force in the horizontal direction is causing this change in motion, find (a) the components of the force and (b) the magnitude of the force. |
| Chapter#6Question#25 | A 30.0-kg girl in a swing is pushed to one side and held at rest by a horizontal force F so that the swing ropes are 30 degrees with respect to the vertical. (a) Calculate the tension in each of the two ropes supporting the swing under these conditions. (b) Calculate the magnitude of F |
| Chapter#6Question#27 | F1 = (-3.00 i + 2.00 j) N<br>F2 = (6.00 i - 4.00 j) N<br>F3 = (2.00 i + 5.00 j) N<br>Acceleration magnitude = 4.23 m/s^2<br><br>a) Find acceleration vector in terms of m<br>b) Find mass |
| | c) Starting from rest, find speed after 5.00 s<br>d) Find velocity components after 5.00 s |
| Chapter#6Question#31 | A 35.0-kg dolphin accelerates opposite to the motion from 12.0 to 7.50 m/s in 2.30 s to join another dolphin in play. What average force was exerted to slow the first dolphin if it was moving horizontally? (The gravitational force is balanced by the buoyant force of the water.) |
| Chapter#6Question#33 | Velocity = (3 i - 2 j) m/s (constant)<br><br>F1 = (3 i + 5 j) N<br>F2 = (4 i - 7 j) N<br><br>Find F3 |

Multimodal Questions:

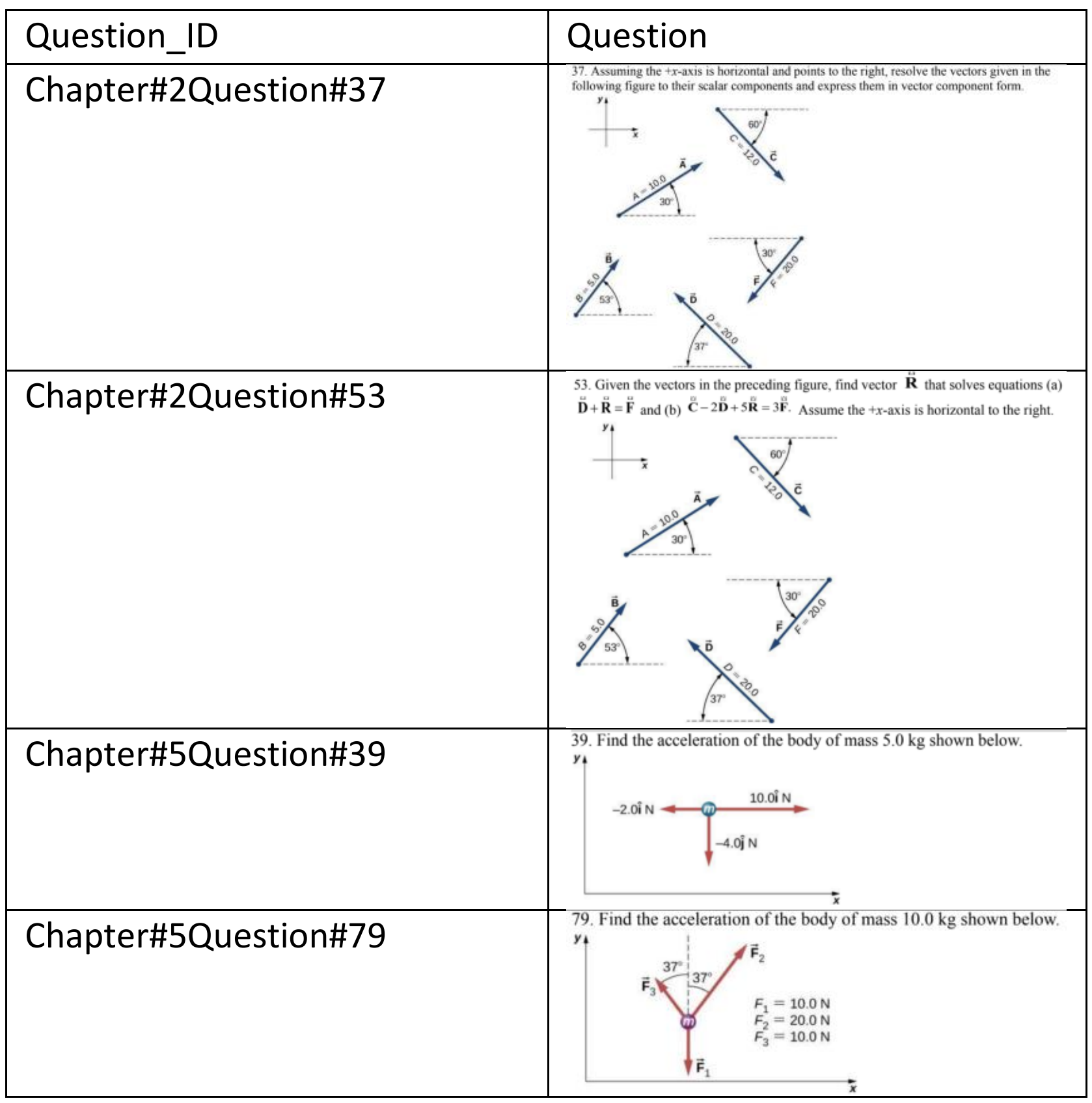

| Question_ID | Question |
|---|---|
| Chapter#2Question#37 | 37. Assuming the +x-axis is horizontal and points to the right, resolve the vectors given in the following figure to their scalar components and express them in vector component form. |
| Chapter#2Question#53 | 53. Given the vectors in the preceding figure, find vector $\vec{R}$ that solves equations (a) $\vec{D}+\vec{R}=\vec{F}$ and (b) $\vec{C}-2\vec{D}+5\vec{R}=3\vec{F}$. Assume the +x-axis is horizontal to the right. |
| Chapter#5Question#39 | 39. Find the acceleration of the body of mass 5.0 kg shown below. |
| Chapter#5Question#79 | 79. Find the acceleration of the body of mass 10.0 kg shown below. |
| Chapter#5Question#81 | 81. Force $\vec{F}_B$ has twice the magnitude of force $\vec{F}_A$. Find the direction in which the particle accelerates in this figure.<br>Page 12 of 16<br>OpenStax *University Physics Volume I*<br>Unit 1: Mechanics<br>Chapter 5: Newton's Laws of Motion |
| Chapter#5Question#87 | 87. As shown below, two identical springs, each with the spring constant 20 N/m, support a 15.0-N weight. (a) What is the tension in spring A? (b) What is the amount of stretch of spring A from the rest position?<br>Page 13 of 16<br>OpenStax *University Physics Volume I*<br>Unit 1: Mechanics<br>Chapter 5: Newton's Laws of Motion |
| Chapter#6#29 | 29. Two muscles in the back of the leg pull upward on the Achilles tendon, as shown below. (These muscles are called the medial and lateral heads of the gastrocnemius muscle.) Find the magnitude and direction of the total force on the Achilles tendon. What type of movement could be caused by this force? |
| Chapter#6#45 | 45. A 2.00 kg block (mass 1) and a 4.00 kg block (mass 2) are connected by a light string as shown; the inclination of the ramp is $40.0°$. Friction is negligible. What is (a) the acceleration of each block and (b) the tension in the string? |
| Chapter#6#61 | 61. Consider the 52.0-kg mountain climber shown below. (a) Find the tension in the rope and the force that the mountain climber must exert with her feet on the vertical rock face to remain stationary. Assume that the force is exerted parallel to her legs. Also, assume negligible force exerted by her arms. (b) What is the minimum coefficient of friction between her shoes and the cliff?<br>Page 9 of 18<br>OpenStax *University Physics Volume I*<br>Unit 1: Mechanics<br>Chapter 6: Applications of Newton's Laws |
| Chapter#6#73 | 73. A child of mass 40.0 kg is in a roller coaster car that travels in a loop of radius 7.00 m. At point A the speed of the car is 10.0 m/s, and at point B, the speed is 10.5 m/s. Assume the child is not holding on and does not wear a seat belt. (a) What is the force of the car seat on the child at point A? (b) What is the force of the car seat on the child at point B? (c) What minimum speed is required to keep the child in his seat at point A? |

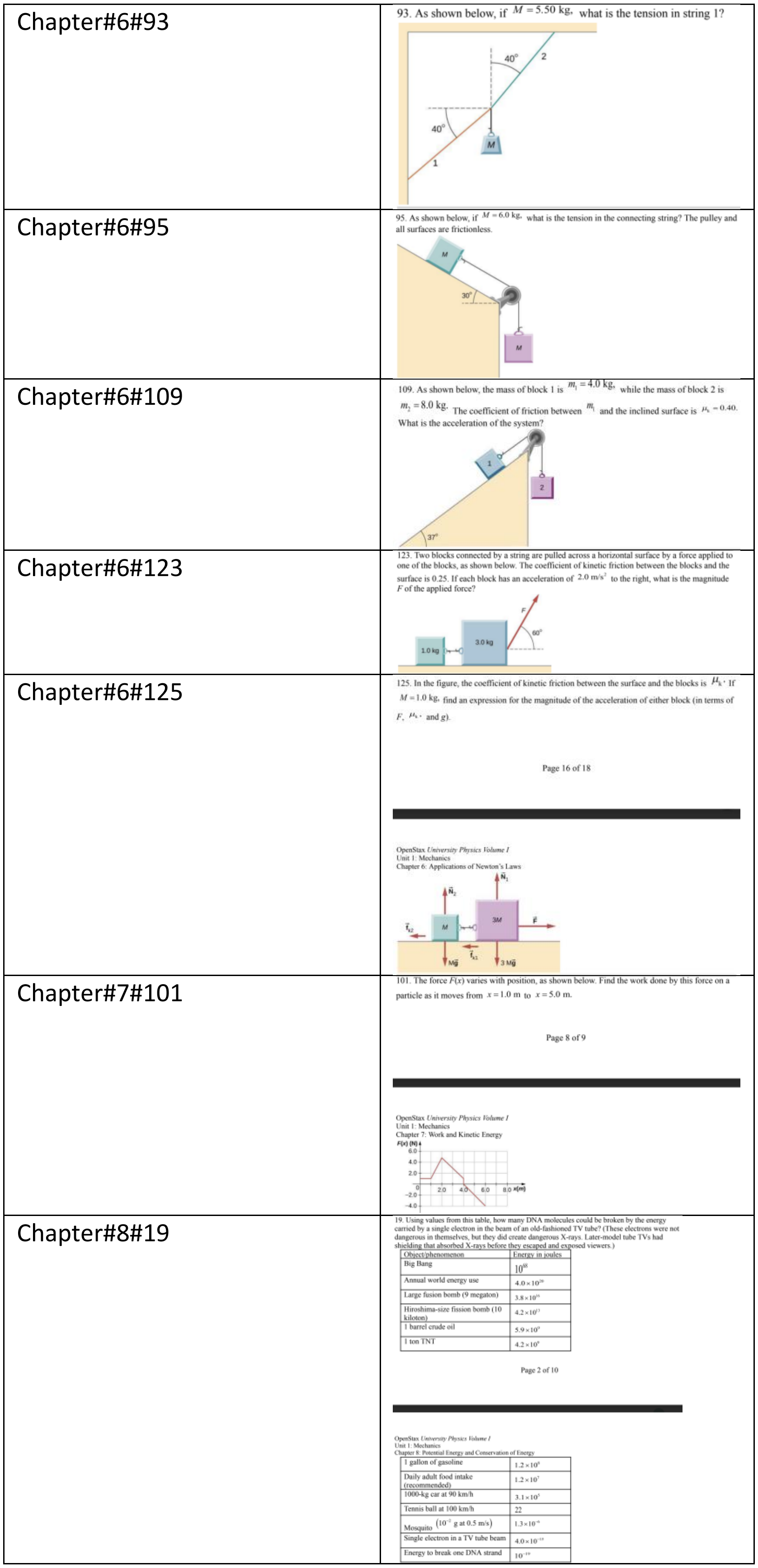

Chapter#6#93

93. As shown below, if $M = 5.50$ kg, what is the tension in string 1?

Chapter#6#95

95. As shown below, if $M = 6.0$ kg, what is the tension in the connecting string? The pulley and all surfaces are frictionless.

Chapter#6#109

109. As shown below, the mass of block 1 is $m_1 = 4.0$ kg, while the mass of block 2 is $m_2 = 8.0$ kg. The coefficient of friction between $m_1$ and the inclined surface is $\mu_k = 0.40$. What is the acceleration of the system?

Chapter#6#123

123. Two blocks connected by a string are pulled across a horizontal surface by a force applied to one of the blocks, as shown below. The coefficient of kinetic friction between the blocks and the surface is 0.25. If each block has an acceleration of $2.0 \text{ m/s}^2$ to the right, what is the magnitude *F* of the applied force?

Chapter#6#125

125. In the figure, the coefficient of kinetic friction between the surface and the blocks is $\mu_k$. If $M = 1.0$ kg, find an expression for the magnitude of the acceleration of either block (in terms of *F*, $\mu_k$, and g).

Page 16 of 18

OpenStax *University Physics Volume I*
Unit 1: Mechanics
Chapter 6: Applications of Newton's Laws

Chapter#7#101

101. The force *F*(*x*) varies with position, as shown below. Find the work done by this force on a particle as it moves from $x = 1.0$ m to $x = 5.0$ m.

Page 8 of 9

OpenStax *University Physics Volume I*
Unit 1: Mechanics
Chapter 7: Work and Kinetic Energy

Chapter#8#19

19. Using values from this table, how many DNA molecules could be broken by the energy carried by a single electron in the beam of an old-fashioned TV tube? (These electrons were not dangerous in themselves, but they did create dangerous X-rays. Later-model tube TVs had shielding that absorbed X-rays before they escaped and exposed viewers.)

| Object/phenomenon | Energy in joules |
|---|---|
| Big Bang | $10^{68}$ |
| Annual world energy use | $4.0 \times 10^{20}$ |
| Large fusion bomb (9 megaton) | $3.8 \times 10^{16}$ |
| Hiroshima-size fission bomb (10 kiloton) | $4.2 \times 10^{13}$ |
| 1 barrel crude oil | $5.9 \times 10^{9}$ |
| 1 ton TNT | $4.2 \times 10^{9}$ |
| 1 gallon of gasoline | $1.2 \times 10^{8}$ |
| Daily adult food intake (recommended) | $1.2 \times 10^{7}$ |
| 1000-kg car at 90 km/h | $3.1 \times 10^{5}$ |
| Tennis ball at 100 km/h | 22 |
| Mosquito $(10^{-2} \text{ g at } 0.5 \text{ m/s})$ | $1.3 \times 10^{-6}$ |
| Single electron in a TV tube beam | $4.0 \times 10^{-15}$ |
| Energy to break one DNA strand | $10^{-19}$ |

Page 2 of 10

OpenStax *University Physics Volume I*
Unit 1: Mechanics
Chapter 8: Potential Energy and Conservation of Energy

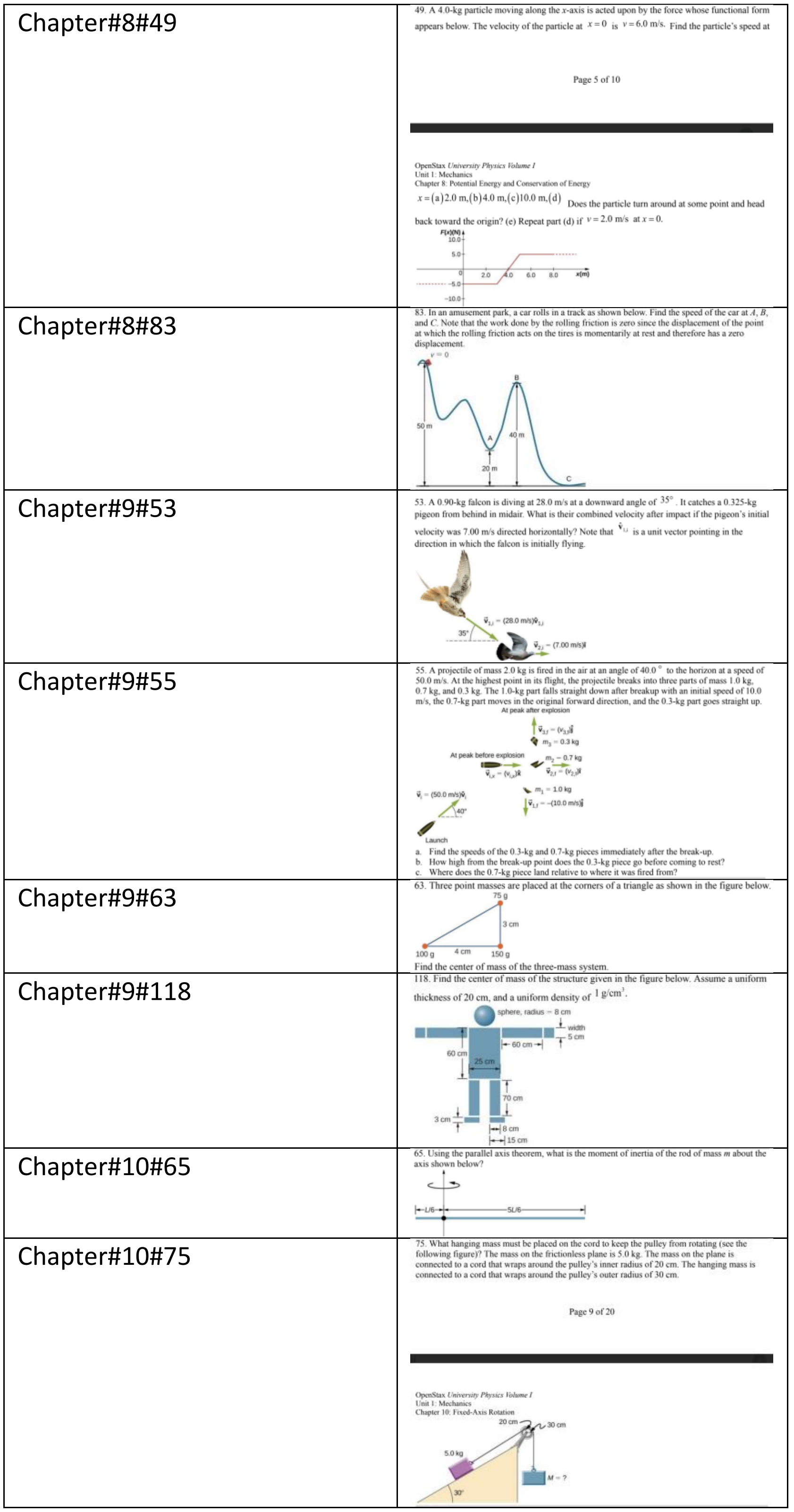

| | |
|---|---|
| Chapter#8#49 | 49. A 4.0-kg particle moving along the $x$-axis is acted upon by the force whose functional form appears below. The velocity of the particle at $x=0$ is $v=6.0$ m/s. Find the particle's speed at <br> Page 5 of 10 <br> OpenStax *University Physics Volume I* <br> Unit 1: Mechanics <br> Chapter 8: Potential Energy and Conservation of Energy <br> $x=(a)2.0$ m, $(b)4.0$ m, $(c)10.0$ m, $(d)$ Does the particle turn around at some point and head back toward the origin? (e) Repeat part (d) if $v=2.0$ m/s at $x=0$. |
| Chapter#8#83 | 83. In an amusement park, a car rolls in a track as shown below. Find the speed of the car at *A*, *B*, and *C*. Note that the work done by the rolling friction is zero since the displacement of the point at which the rolling friction acts on the tires is momentarily at rest and therefore has a zero displacement. |
| Chapter#9#53 | 53. A 0.90-kg falcon is diving at 28.0 m/s at a downward angle of $35°$. It catches a 0.325-kg pigeon from behind in midair. What is their combined velocity after impact if the pigeon's initial velocity was 7.00 m/s directed horizontally? Note that $\hat{v}_{1,i}$ is a unit vector pointing in the direction in which the falcon is initially flying. |
| Chapter#9#55 | 55. A projectile of mass 2.0 kg is fired in the air at an angle of 40.0° to the horizon at a speed of 50.0 m/s. At the highest point in its flight, the projectile breaks into three parts of mass 1.0 kg, 0.7 kg, and 0.3 kg. The 1.0-kg part falls straight down after breakup with an initial speed of 10.0 m/s, the 0.7-kg part moves in the original forward direction, and the 0.3-kg part goes straight up. <br> a. Find the speeds of the 0.3-kg and 0.7-kg pieces immediately after the break-up. <br> b. How high from the break-up point does the 0.3-kg piece go before coming to rest? <br> c. Where does the 0.7-kg piece land relative to where it was fired from? |
| Chapter#9#63 | 63. Three point masses are placed at the corners of a triangle as shown in the figure below. <br> Find the center of mass of the three-mass system. |
| Chapter#9#118 | 118. Find the center of mass of the structure given in the figure below. Assume a uniform thickness of 20 cm, and a uniform density of $1\ g/cm^3$. |
| Chapter#10#65 | 65. Using the parallel axis theorem, what is the moment of inertia of the rod of mass $m$ about the axis shown below? |
| Chapter#10#75 | 75. What hanging mass must be placed on the cord to keep the pulley from rotating (see the following figure)? The mass on the frictionless plane is 5.0 kg. The mass on the plane is connected to a cord that wraps around the pulley's inner radius of 20 cm. The hanging mass is connected to a cord that wraps around the pulley's outer radius of 30 cm. <br> Page 9 of 20 <br> OpenStax *University Physics Volume I* <br> Unit 1: Mechanics <br> Chapter 10: Fixed-Axis Rotation |
| Chapter#10#77 | 77. Calculate the torque about the z-axis that is out of the page at the origin in the following figure, given that $F_1 = 3\text{ N}$, $F_2 = 2\text{ N}$, $F_3 = 3\text{ N}$, $F_4 = 1.8\text{ N}$ |
| Chapter#10#95 | 95. A uniform rod of mass and length is held vertically by two strings of negligible mass, as shown below. (a) Immediately after the string is cut, what is the linear acceleration of the free end of the stick? (b) Of the middle of the stick? |
| Chapter#10#115 | 115. A pendulum consists of a rod of length 2 m and mass 3 kg with a solid sphere of mass 1 kg and radius 0.3 m attached at one end. The axis of rotation is as shown below. What is the angular velocity of the pendulum at its lowest point if it is released from rest at an angle of $30°$?<br>Page 16 of 20<br>OpenStax *University Physics Volume I*<br>Unit 1: Mechanics<br>Chapter 10: Fixed-Axis Rotation |
| Chapter#10#123 | 123. A disk of mass *m*, radius *R*, and area *A* has a surface mass density $\sigma = \frac{mr}{AR}$ (see the following figure). What is the moment of inertia of the disk about an axis through the center? |
| Chapter#11#65 | 65. Twin skaters approach one another as shown below and lock hands. (a) Calculate their final angular velocity, given each had an initial speed of 2.50 m/s relative to the ice. Each has a mass of 70.0 kg, and each has a center of mass located 0.800 m from their locked hands. You may approximate their moments of inertia to be that of point masses at this radius. (b) Compare the initial kinetic energy and final kinetic energy. |
| Chapter#11#77 | 77. The axis of Earth makes a 23.5° angle with a direction perpendicular to the plane of Earth's orbit. As shown below, this axis precesses, making one complete rotation in 25,780 y.<br>(a) Calculate the change in angular momentum in half this time.<br>(b) What is the average torque producing this change in angular momentum?<br>(c) If this torque were created by a pair of forces acting at the most effective point on the equator, what would the magnitude of each force be? |
| Chapter#12#27 | 27. What force must be applied at point *P* to keep the structure shown in equilibrium? The weight of the structure is negligible. |
| Chapter#12#31 | 31. The uniform seesaw is balanced at its center of mass, as seen below. The smaller boy on the right has a mass of 40.0 kg. What is the mass of his friend? |
| Chapter#12#33 | 33. The uniform seesaw shown below is balanced on a fulcrum located 3.0 m from the left end. The smaller boy on the right has a mass of 40 kg and the bigger boy on the left has a mass 80 kg. What is the mass of the board? |
| Chapter#12#39 | 39. The forearm shown below is positioned at an angle $\theta$ with respect to the upper arm, and a 5.0-kg mass is held in the hand. The total mass of the forearm and hand is 3.0 kg, and their center of mass is 15.0 cm from the elbow. (a) What is the magnitude of the force that the biceps muscle exerts on the forearm for $\theta = 60^\circ$? (b) What is the magnitude of the force on the elbow joint for the same angle? (c) How do these forces depend on the angle $\theta$? |
| Chapter#12#41 | 41. The uniform boom shown below weighs 700 N, and the object hanging from its right end weighs 400 N. The boom is supported by a light cable and by a hinge at the wall. Calculate the tension in the cable and the force on the hinge on the boom. Does the force on the hinge act along the boom?<br>Page 4 of 11<br>OpenStax *University Physics Volume I*<br>Unit 1: Mechanics<br>Chapter 12: Static Equilibrium and Elasticity |
| Chapter#12#71 | 71. A uniform 4.0-m plank weighing 200.0 N rests against the corner of a wall, as shown below. There is no friction at the point where the plank meets the corner. (a) Find the forces that the corner and the floor exert on the plank. (b) What is the minimum coefficient of static friction between the floor and the plank to prevent the plank from slipping? |
| Chapter#12#75 | 75. A horizontal force $\vec{F}$ is applied to a uniform sphere in direction exact toward the center of the sphere, as shown below. Find the magnitude of this force so that the sphere remains in static equilibrium. What is the frictional force of the incline on the sphere? |
| Chapter#12#77 | 77. Two wheels *A* and *B* with weights *w* and 2*w*, respectively, are connected by a uniform rod with weight *w*/2, as shown below. The wheels are free to roll on the sloped surfaces. Determine the angle that the rod forms with the horizontal when the system is in equilibrium. *Hint:* There are five forces acting on the rod, which is two weights of the wheels, two normal reaction forces at points where the wheels make contacts with the wedge, and the weight of the rod. |
| Chapter#12#79 | 79. In order to lift a shovelful of dirt, a gardener pushes downward on the end of the shovel and pulls upward at distance $l_2$ from the end, as shown below. The weight of the shovel is $m\vec{g}$ and acts at the point of application of $\vec{F}_2$. Calculate the magnitudes of the forces $\vec{F}_1$ and $\vec{F}_2$ as functions of $l_1$, $l_2$, *mg*, and the weight *W* of the load. Why do your answers not depend on the angle $\theta$ that the shovel makes with the horizontal?<br>Page 9 of 11<br>OpenStax *University Physics Volume I*<br>Unit 1: Mechanics<br>Chapter 12: Static Equilibrium and Elasticity |
| Chapter#12#81 | 81. The pole shown below is at a $90.0^\circ$ bend in a power line and is therefore subjected to more shear force than poles in straight parts of the line. The tension in each line is $4.00 \times 10^4\ \text{N}$, at the angles shown. The pole is 15.0 m tall, has an 18.0 cm diameter, and can be considered to have half the strength of hardwood. (a) Calculate the compression of the pole. (b) Find how much it bends and in what direction. (c) Find the tension in a guy wire used to keep the pole straight if it is attached to the top of the pole at an angle of $30.0^\circ$ with the vertical. The guy wire is in the opposite direction of the bend. |
| Chapter#14#89 | 89. A container of water has a cross-sectional area of $A = 0.1\ \text{m}^2$. A piston sits on top of the water (see the following figure). There is a spout located 0.15 m from the bottom of the tank, open to the atmosphere, and a stream of water exits the spout. The cross sectional area of the spout is $A_s = 7.0 \times 10^{-4}\ \text{m}^2$. (a) What is the velocity of the water as it leaves the spout? (b) If the opening of the spout is located 1.5 m above the ground, how far from the spout does the water hit the floor? Ignore all friction and dissipative forces.<br>20 kg<br>0.5 m<br>$\rho = 1000 \frac{\text{kg}}{\text{m}^3}$<br>0.15 m |
| Chapter#17#41 | 41. Consider the graph shown below of a compression wave. Shown are snapshots of the wave function for $t = 0.000\ \text{s}$ (blue) and $t = 0.005\ \text{s}$ (orange). What are the wavelength, maximum displacement, velocity, and period of the compression wave?<br>s(x, t = 0.005s)<br>s(x, t = 0s)<br>y(mm)<br>3 2 1 0 −1 −2 −3<br>0 1 2 3 4 5 6 7 8 9 10 11 12 13 14 15<br>x(m) |
| Chapter#17#53 | 53. Ultrasonic sound waves are often used in methods of nondestructive testing. For example, this method can be used to find structural faults in a steel I-beams used in building. Consider a 10.00 meter long, steel I-beam with a cross-section shown below. The weight of the I-beam is 3846.50 N. What would be the speed of sound through in the I-beam?<br>$(Y_{steel} = 200\ \text{GPa}, \beta_{steel} = 159\ \text{GPa})$<br><br>Page 6 of 19<br><br>OpenStax *University Physics Volume I*<br>Unit 2: Waves and Acoustics<br>Chapter 17: Sound<br>5.0 cm<br>2.5 cm<br>10.00 cm<br>2.5 cm |

## Appendix B: Correctness Verification Using an LLM-Judge

To improve evaluation efficiency, the LLM-judge pipeline was optimized using asynchronous API execution. Judge evaluations across multiple models were executed in parallel, significantly reducing end-to-end latency.

This optimization resulted in a 72% reduction in per-response evaluation time, enabling scalable assessment of large problem sets without compromising evaluation fidelity.

### Judge Design

An independent LLM-based evaluator was developed to assess correctness. To ensure robustness against model-specific biases, we employed a multi-judge framework consisting of Gemini-3 Pro Preview, GPT-5 (high-effort reasoning), and Sonnet-4.5.

Each judge independently evaluated the model response against the ground truth using an identical verification prompt and semantic equivalence rules. Final correctness was determined via majority voting across the three judges.

This design mitigates single-model evaluation bias and improves reliability of correctness assessment, particularly in edge cases involving ambiguous representations or multi-step reasoning.

### Model Configurations

1. Gemini
    1. model="gemini-3-pro-preview"
    2. thinking_level="HIGH"
    3. max_output_tokens=7000
2. OpenAI
    1. model="gpt-5"
    2. Reasoning effort = "high"
    3. Text verbosity  = "low"
3. Claude Sonnet
    1. CLAUDE_MODEL = "claude-sonnet-4-5-20250929"
    2. Thinking budget_tokens = 6000
    3. max_tokens=7000

### Judge Inputs

1. Ground-truth answer image
2. Model response (text)

Judge Procedure

1. Extract ground-truth answers directly from the image.
2. Identify all subparts of the question.
3. Normalize both ground-truth and model answers into a canonical representation.
4. Apply semantic equivalence rules:

a. Algebraic equivalence

b. Vector representation equivalence

c. Reference-frame consistency

d. Apply numeric tolerance (≈1% relative difference).

5. Each judge produces an independent binary verdict (Correct or Wrong) without explanatory text. The final system verdict is computed via majority voting across all judges. In cases of disagreement, the majority decision is taken as the final correctness label.
6. The full judge logic was enforced via a fixed system prompt:

You are an answer verification system.

Your task is to compare the ground truth answer extracted from the image with the model answer provided in the JSON and return a binary verdict.

Core Instructions:

1. Extract the ground truth answer exactly as shown in the image.

2. Read the model's answer from the provided JSON.

3. Compare all subparts (if the question has multiple parts).

4. If even one subpart is incorrect, incomplete, or missing, the verdict MUST be Wrong.

5. Do not infer intent or give partial credit.

Semantic Equivalence Requirement (MANDATORY):

Before deciding Correct or Wrong, you MUST check whether the ground truth and model answer are semantically equivalent even if expressed in different valid forms. You MUST normalize both answers into a common canonical representation before comparison.

Acceptable equivalence transformations include (when applicable):

1. Vector representations: Cartesian components (ai + bj), magnitude–direction form, unit-vector form, or polar vs Cartesian. Convert one representation into the other and compare consistently.
2. Reference frame differences: If each answer explicitly states a different origin or reference point, transform coordinates into the same reference frame before comparison.
3. Algebraic equivalence: Simplified vs unsimplified expressions, factored vs expanded forms, exact vs approximate values (e.g., sqrt(2) vs 1.414).
4. Derived equivalence: If one answer gives magnitude and direction, compute the implied components (or vice versa) and compare.
5. Numeric Tolerance Rule (apply AFTER canonicalization):

a. Minor numerical differences caused by rounding or approximation MUST be treated as Correct.

b. Use a reasonable tolerance (approximately 1% relative difference or small absolute error), unless the problem explicitly requires exact precision. Examples of acceptable matches: 5 vs 4.99, 3.28 vs 3.29, 25 vs 24.85.

6. Disallowed Equivalence: Do NOT mark Correct if equivalence would require changing physical assumptions, ignoring stated reference frames, flipping axes without justification, or inventing unstated transformations.
7. Output Rules:

a. Output exactly one line.

b. Use the following format verbatim: verdict: <Correct|Wrong>, ground_truth: <value>, model_answer: <value>

c. If there are multiple subparts, list them as comma-separated pairs.

d. Only output the final line. Do not include explanations, reasoning, or extra text. Example: verdict: Correct, ground_truth: a=3, b=7, model_answer: a=2.98, b=6.99

<u>Tools Used</u>

1. Python

2. Google Colab
3. APIs:
    1. Google Generative AI API (Gemini-3 Pro Preview)
    2. OpenAI API (GPT-5,)
    3. Anthropic API (Claude Sonnet-4.5)